\documentclass{appolb}
\usepackage{graphicx}
\usepackage{cite}
\usepackage{dsfont}
\usepackage{physics}
\usepackage{hyperref}
\usepackage{amsmath}
\usepackage{amssymb}
\usepackage{ulem}
\usepackage{color}                                                     %%   change color%------------------------------------------------------
\usepackage{xcolor}

\begin{document}
\title{\color{blue} Facets of Many Body Localization%
    \thanks{Presented at  Concepts in Strongly Correlated Quantum Matter 25}%
}
\author{Konrad Pawlik, Maksym Prodius, Pedro R. Nic\'acio Falc\~ao
\address{Szkoła Doktorska Nauk \'Scis\l{}ych i Przyrodniczych, Uniwersytet Jagiello\'nski,  \L{}ojasiewicza 11, PL-30-348 Krak\'ow, Poland}
\\[3mm]
{Jakub Zakrzewski % of different affiliation
\address{Instytut Fizyki Teoretycznej, Wydzia\l{} Fizyki, Astronomii i Informatyki Stosowanej,
    Uniwersytet Jagiello\'nski,  \L{}ojasiewicza 11, PL-30-348 Krak\'ow, Poland}
\address{Mark Kac Complex Systems Research Center, Uniwersytet Jagiello{\'n}ski, Krak{\'o}w, Poland }
}
}
\maketitle
\begin{abstract}
    Many-body localization (MBL) appears to be a robust example of ergodicity breaking in many-body interacting systems. Here, we review different aspects of MBL,
    concentrating on various ways the disorder may be introduced into the system studied. In particular, we consider both the random and quasiperiodic diagonal
    (i.e., on-site) disorder as well as bond disorder as realized in randomly distributed atoms interacting via long-range interactions. We also review the quantum sun model, which seems to be the ideal, albeit artificial, model exhibiting MBL.
\end{abstract}

\section{Introduction}
The foundations of classical statistical mechanics rest on the ergodic hypothesis, which states that over long periods, a system explores all accessible microstates with equal probability, effectively erasing the memory of its initial conditions. Generalizing this concept to the quantum realm is nontrivial, as phase-space trajectories are not well defined. Furthermore, isolated quantum systems undergo unitary evolution, with a time-invariant probability of finding the system in a specific Hamiltonian eigenstate. This contrasts sharply with classical systems, which explore different states during the evolution. These challenges motivated the formulation of the Eigenstate Thermalization Hypothesis (ETH)~\cite{Deutsch91,Srednicki94}, which posits that, in ergodic systems, the expectation values of observables within individual eigenstates match those of the microcanonical ensemble at corresponding eigenenergies. In this framework, even closed quantum systems can be ergodic, with subsystems perceiving the remainder of the system as an effective thermal bath.

In contrast to classical physics, the paradigm of quantum thermalization is not universal. The most robust exception arises in the presence of strong disorder, leading to the phenomenon of many-body localization (MBL). Historically rooted in Anderson localization~\cite{Anderson58}, where the transport of non-interacting particles vanishes due to the presence of a random potential, MBL generalizes this phenomenon to systems where particles are allowed to interact~\cite{Gornyi05, Basko06}. As a dynamical phase of matter, it is characterized by a persistent memory of the initial state during time evolution~\cite{Serbyn13b, Huse14, Schreiber15, Smith16}, the suppression of transport~\cite{Nandkishore15, Znidaric16}, a slow entanglement growth~\cite{Znidaric08, Bardarson12,Serbyn13a}, and Poisson spectral statistics~\cite{Oganesyan07}, typical of integrable systems.

This overview addresses the multifaceted nature of MBL systems. We move beyond the paradigmatic uniform random box distribution of diagonal disorder, covered in Sec.~\ref{sec:mbl}, to explore distinct mechanisms that induce localization. We begin in Sec.~\ref{sec:qp} by considering a quasiperiodic potential, which naturally emerges in experimental realizations by imposing a standing wave with a period incommensurate with the main lattice~\cite{Schreiber15, Luschen17,Lukin19, Rispoli19, Leonard23}. This experimental advantage, however, comes at the cost of introducing correlations between different on-site potentials. In Sec.~\ref{sec:bd}, we discuss bond-disordered models. This scenario is particularly relevant to experimental platforms utilizing Rydberg atoms or dipolar gases, where the random spatial configuration of particles manifests as disordered interaction strengths. Next, in Sec.~\ref{sec:qs}, we present the Quantum Sun model, a toy model of a localized chain interacting with a thermal bath. This model exhibits localization that is significantly more robust than in other MBL systems, thereby avoiding many of their inherent complications discussed in this review. Finally, in Sec.~\ref{sec:con}, we provide a summary and outlook.

\section{``Standard MBL''\label{sec:mbl}}

In 1D spin-$\tfrac{1}{2}$ chains,  models of MBL are commonly defined by the Hamiltonian of the form
\begin{equation}
    \hat{H} = \hat{H_0} + \sum_{j=1}^L h_j \hat{S}^z_j,
    \label{eq:standard_MBL}
\end{equation}
where $H_0$ is a disorder-free, short-range interacting Hamiltonian, $\hat{S}^{\alpha}_j$ ($\alpha\in\{x,y,z\}$) are spin operators, and $h_j$ is a disordered on-site field drawn from a uniform distribution with amplitude $W$, i.e., $h_j \in \left[-W, W\right]$. Among different models studied in this context, the XXZ spin chain has received significant attention over the years, being now considered the paradigmatic model for MBL~\cite{Pal10,Abanin19,Sierant25}. Its Hamiltonian reads
\begin{equation}
    \hat{H}_{0} = \sum_{j=1}^{L} J \big(\hat{S}_{j}^{x}\hat{S}_{j+1}^{x} + \hat{S}_{j}^{y}\hat{S}_{j+1}^{y} +\Delta\hat{S}_{j}^{z}\hat{S}_{j+1}^{z} \big),
    \label{Hamilt_XXZ}
\end{equation}
where $\Delta$ is the interaction strength, and the overall energy scale is set as $J=1$. In the non-interacting limit ($\Delta=0$), the model is localized for any $W>0$, with all eigenstates being exponentially localized~\cite{Anderson58, Evers00}. Upon turning on the interactions, the system quickly becomes ergodic for a small but finite disorder strength. In this regime, the eigenstates exhibit volume-law entanglement entropy and Wigner-Dyson level-spacing statistics~\cite{Dalessio16}, while local observables follow ETH and any memory of a typical initial state is washed away.

As the disorder strength $W$ increases, the system undergoes a qualitative change in all of the previously mentioned characteristics. The average gap ratio~\cite{Oganesyan07}, defined as
\begin{equation}
    \langle r \rangle = \Big\langle \frac{\min(\delta_n, \delta_{n+1})}{\max(\delta_n, \delta_{n-1})}\Big \rangle, \quad \delta_n = E_{n} - E_{n-1},
\end{equation}
with $\langle \cdot \rangle$ denoting the average over disorder realizations and energy range of interest, undergoes a transition from GOE ($\langle r\rangle \approx 0.5307$) statistics in the ergodic regime to Poisson ($\langle r\rangle \approx 0.386$) statistics in the localized regime. A similar crossover happens for the bipartite entanglement entropy of eigenstates, $ \mathcal{S} = -\mathrm{Tr}(\rho_A \log \rho_A)$ (where $\rho_A$ is a partial density matrix obtained by tracing over the $B$ subsystem, assuming a bipartite system's splitting into $A$ and $B$), which changes from volume-law to area-law scaling. Therefore, these quantities serve as standard indicators of ergodicity breaking.

A major conceptual advancement in the understanding of MBL was the realization that the system can be described by an extensive set of quasi-local integrals of motion~\cite{Serbyn13b, Huse14}. This explains key hallmarks of MBL: {the logarithmic growth of entanglement entropy during the time evolution of low-entropy initial states, and the power-law temporal decay of local observable fluctuations toward stationary initial state dependent values~\cite{Serbyn14}}.

\begin{figure}[t!]
    \centerline{%
        \includegraphics[width=1.0\linewidth]{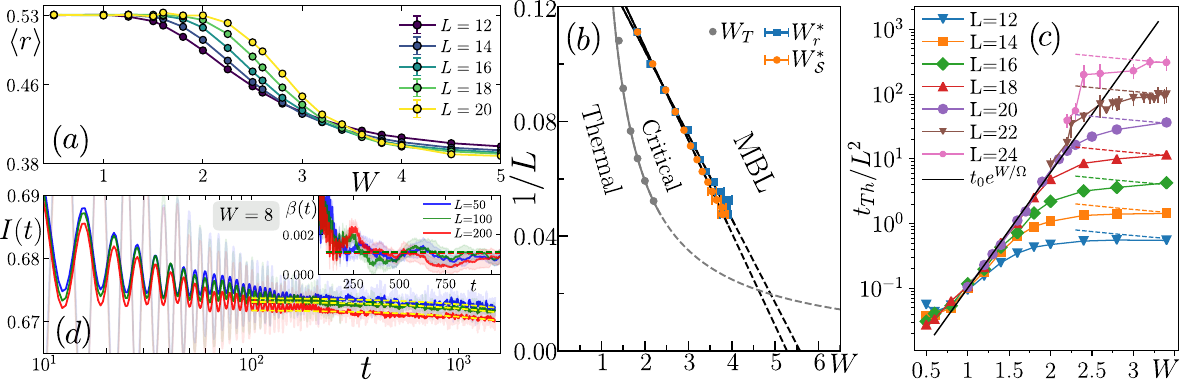}}
    \caption{{\it The paradigmatic XXZ model with $\Delta=1$}. (a) Average gap ratio $\langle r\rangle$ as a function of the disorder strength $W$. (b) Estimate of the critical point depending on the method used scales as $1/L$ or as $L$ -- adapted from \cite{Sierant20p}, @(2020) the American Physical Society. (c) Scaling of the Thouless time $t_{\rm Th}$, black solid line corresponding to scaling~\eqref{eq:thouless} -- adapted from \cite{Sierant20b}, @(2020) the American Physical Society. (d) Imbalance dynamics of the Neel state for $W=8$, inset showing exponent of the effective power-law decay -- adapted from \cite{Sierant22} @(2022) the American Physical Society. }
    \label{fig:mbl}
\end{figure}

Despite this progress, the status of MBL remains an open problem. Unlike conventional phase transitions, the ETH-MBL crossover exhibits pronounced finite-size drifts, which impose a significant threat to the stability of MBL in the thermodynamic limit. In Fig.~\ref{fig:mbl}, we illustrate several of these issues: In standard ergodicity breaking indicators, such as the gap ratio $\langle r \rangle$ shown in Fig.~\ref{fig:mbl}(a), the ergodic boundary shows a persistent linear drift with increasing system size, which may be indicative of eventual thermalization. The crossing between two different curves also shifts with increasing $L$, albeit at a slower rate. The coexistence of two incompatible scalings suggests that a consistent interpretation would require identifying a critical system size at which one trend breaks down.
The extrapolation of both behaviors suggests a critical system size of $L\approx50$, far beyond state-of-the-art methods~\cite{Sierant20p}. Furthermore, the extrapolation of the crossing points indicates a critical disorder strength of $W\approx 5$~\cite{Sierant20p}, substantially larger than earlier estimations~\cite{Luitz15}. Such a result agrees with recent scaling theories of many-body localization~\cite{niedda25r}, although other studies suggest that interaction-driven instabilities and system-wide resonances can shift the critical point much further~\cite{Morningstar22,Laflorencie25}.

The study of the Thouless time $t_{\rm Th}$ - namely the time after which the dynamics of the system are universal and governed by the respective random matrix ensemble - demonstrates the difficulty caused by finite-size effects when concluding on the thermodynamic limit of many-body models. It was observed~\cite{Suntajs20e}, that the numerical scaling of the Thouless time in disordered quantum spin chains agrees well with
\begin{equation}
    t_{\rm Th}= t_0 e^{W/\Omega}L^2,\label{eq:thouless}
\end{equation}
where constants $t_0, \Omega$ depend on the microscopic details of the model. If the scaling were to continue up to the thermodynamic limit, it would imply that the Thouless time would eventually become negligible compared to the Heisenberg time $t_{\rm H}$, which scales exponentially with system size and represents the timescale after which the discreteness of the spectrum manifests itself. This would imply ergodic behavior; thus, MBL would simply be a finite-size effect that vanishes when the system size is large enough. It was later discovered~\cite{Sierant20b} that 3D and 5D Anderson models, known to have a localized phase in the thermodynamic limit, share a scaling similar to Eq.~\eqref{eq:thouless}. Due to the lack of interactions, the position of the critical point can still be determined in this case, provided one knows the exponent governing subdiffusion at the Anderson transition; however, for the interacting models, no such procedure exists. Nonetheless, the similarity to the non-interacting case suggested posing an alternative hypothesis: that the scaling~\eqref{eq:thouless} is valid only in a limited range of system sizes, and its breakdown signals a slowdown of transport as the MBL transition is approached. The breakdown of the scaling is demonstrated in the random XXZ model, see Fig.~\ref{fig:mbl}c), where the largest system sizes available to the state-of-the-art numerics violate it.

In time-dynamics results, signatures of ergodicity are also manifested. In Fig.~\ref{fig:mbl}(d), we show the time evolution of the spin-correlation function $I(t) = c_0^{-1}\sum_{i=1}^{L}\langle S_{i}^{z}(t)S_i^{z}(0)\rangle$, where $c_0$ ensures normalization $I(0)=1$. In an ideal localized phase, the curves for different system sizes would flatten out, and no dynamical behavior would be observed. However, in the XXZ model, we observe a consistent power-law decay in time, as shown in the main panel. Although the extracted power-law exponent is small, as shown in the inset of Fig.~\ref{fig:mbl}(d) (the inset shows the time-averaged exponent of the approximate power law decay, for details see \cite{Sierant22}), it indicates that MBL may ultimately be unstable in the thermodynamic limit~\cite{Sels20, Sierant22, Evers23}.

\section{Quasiperiodic potential\label{sec:qp}}

Another way to break the translational invariance of the system is to replace the uncorrelated disordered fields with a quasiperiodic (QP) potential,
\begin{equation}
    \hat{H} = \hat{H_0} + \sum_{j=1}^L W\cos(2\pi \kappa j +\varphi) \hat{S}^z_j,
    \label{eq:QP}
\end{equation}
where $\kappa$ is an incommensurability factor, typically chosen as $\kappa=(\sqrt{5}-1)/2$, and $\varphi$ is a global phase uniformly sampled from the interval $[0,2\pi]$ for each realization. In the non-interacting limit ($\Delta=0$), the localization properties of the XX model differ qualitatively from the random disorder case~\cite{Aubry80}. In particular, the eigenstates undergo a sharp extended-localized transition in one-dimensional systems, with a critical disorder strength at $W=1$ (in the units used here), rather than being localized for any $W>0$.

In the interacting {XXZ chain}, this problem was first addressed in~\cite{Iyer13}, which showed that such a {QP} potential can also host an MBL transition upon increasing $W$. A key conceptual difference is that the correlated structure of the QP potential suppresses rare regions of anomalously weak disorder, known as Griffiths regions~\cite{Vojta10}, which may lead to avalanche mechanisms that destabilize MBL~\cite{Gopalakrishnan16,Gopalakrishnan15,Agarwal17,DeRoeck17,Luitz17,Peacock23,Szoldra24}. As mentioned above, the QP potential is a favorite of many experiments in optical lattice settings \cite{Schreiber15,Lukin19,Rispoli19,Leonard23}.

Nevertheless, standard ergodicity indicators such as the average gap ratio and the half-chain entanglement entropy appear similar to the random disordered case. A clear difference only emerges upon further inspection, with sample-to-sample fluctuations of these observables behaving markedly differently~\cite {Khemani17,Sierant19b}. The finite-size effects are also weaker in QP~\cite{Aramthottil21,Falcao24}, with the ergodic boundary drifting logarithmically with $L$. Differences are also observed in dynamical properties, with a faster suppression of transport in the QP case~\cite{Prelovsek23} and persistent oscillations in the imbalance dynamics, again supporting stability of the MBL phase~\cite{Sierant22}. Recent experiments in 2D ultracold atoms also show much weaker finite-size effects in QP spin chains~\cite{hur2025stability}.

\begin{figure}[t!]
    \centering
    \includegraphics[width=1\columnwidth]{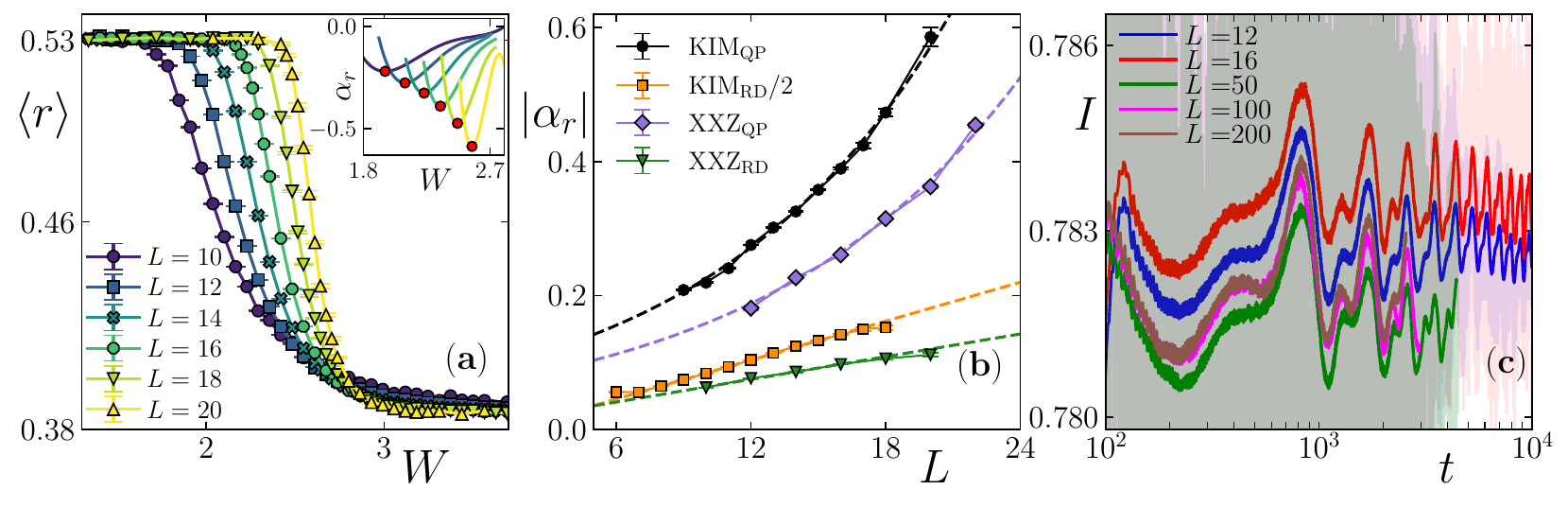}
    \caption{{\it Key features of MBL with QP potential}: (a) Average gap ratio $\langle r \rangle$ as a function of disorder strength $W$ in the Kicked Ising model. The inset shows the derivatives across the transition, with red dots marking the maximum (in absolute value) of the derivatives $\alpha_r$. (b) This maximum scales exponentially with the system size in QP potential as opposed to a linear growth in the random potential -- adapted from \cite{Falcao24}, @(2024) the American Physical Society. (c)~Imbalance in the XXZ model in QP potential reveals strong similarities between different system sizes suggesting MBL
        -- adapted from \cite{Sierant22}, @(2024) the American Physical Society.}
    \label{Fig:Summ_QP}
\end{figure}

In Fig.~\ref{Fig:Summ_QP}, we highlight key differences between RD systems and QP potential. As shown in Fig.~\ref{Fig:Summ_QP}(a), the ETH-MBL crossover in the average gap ratio $\langle r \rangle$ appears similar to the RD case. However, the ergodic boundary does not drift linearly with $L$, indicating a possible stability of MBL~\cite{Aramthottil21}. As shown in the inset and in Fig.~\ref{Fig:Summ_QP}(b), the derivatives $\alpha_r=dr/dW$ sharpen up very quickly with increasing $L$, with $|\alpha_r|$ flowing exponentially toward a non-analytical behavior in QP models (either for the XXZ model or the Kicked Ising model), in contrast with the linear drift in the RD case~\cite{Falcao24}. The kicked Ising model (KIM) is defined via a periodic unitary evolution of the form~\cite{Prosen02}
\begin{equation}
    \hat{U}_{\mathrm{KIM}} = e^{-i \sum_{j=1}^L g \sigma_{j}^x}e^{-i\left( \sum_{j=1}^{L-1} J \sigma_j^{z}\sigma_{j+1}^{z} + \sum_{j=1}^{L} h_j\sigma_j^{z} \right) },
    \label{KIM}
\end{equation}
where $h_j$ is the on-site random magnetic field. KIM shows milder finite-size effects in random disorder \cite{Sierant23f}.
Further differences appear in the imbalance dynamics of an initial N\'eel state, where robust oscillations persist with increasing system size $L$ [Fig~\ref{Fig:Summ_QP}(c)], providing additional evidence for a stable MBL phase in quasiperiodic spin chains~\cite{Sierant22}.

\section{Bond-disordered models\label{sec:bd}}

Interestingly, there exists another route to introducing randomness in lattice many-body systems. Instead of random (or quasiperiodic) onsite fields, one can randomize the couplings themselves, such as the tunneling amplitudes and/or interaction strengths. Experimentally, this can be realized using Rydberg atoms arranged in tweezer arrays \cite{Signoles21}. Rydberg atoms are naturally characterized by strong long-range dipolar interactions, and when placed in optical tweezer traps in the regime of Rydberg blockade, they effectively realize long-range interacting lattice spin-$1/2$ models \cite{Browaeys20}. Since the dipolar couplings depend on the distance between atoms, disorder can be introduced in the form of random atomic positions. Such Hamiltonians can be expressed as:
\begin{equation}
    \hat{H} = \sum_{i \neq j}^{L} J_{i, j} \left(\hat{S}^x_i \hat{S}^x_j + \hat{S}^y_i \hat{S}^y_j \right) + \sum_{i \neq j}^{L} \Delta_{i, j} \hat{S}^z_i \hat{S}^z_j,
    \label{eq:bond_disorder_hamiltonian}
\end{equation}
where $J_{i,j} = \frac{J}{|x_i - x_j|^n}$ and $\Delta_{i,j} = \frac{\Delta}{|x_i - x_j|^m}$. Here, $x_i$ denotes the position of the $i$'s atom, $J$ and $\Delta$ are distance-independent constants, and the exponents $n$ and $m$ are typically $3$ or $6$, depending on the interaction regime in which the Rydberg atoms are tuned.

\begin{figure}[t!]
    \centering
    \includegraphics[width=1.0\linewidth]{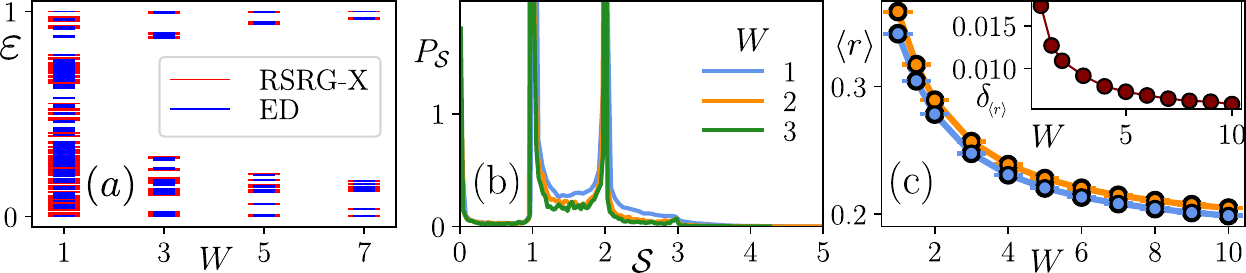}
    \caption{{\it Localization properties of the bond-disordered model} defined by Hamiltonian (\ref{eq:bond_disorder_hamiltonian}) with $m = n = 6$, $J = 1$ and $\Delta = -0.73$. (a)  Comparison of the rescaled energies $\varepsilon$ computed from ED and obtained from RSRG-X for a one disorder realization. (b) Normalized distributions of the half-chain entanglement entropy for $L=20$ and several values of $W$. (c) Mean gap ratio as a function of disorder strength $W$. The blue curve shows the RSRG-X predictions ($\expval{r}_{RG}$), while orange corresponds to the ED results ($\expval{r}_{ED}$). The difference between two approaches $\delta_{\expval{r}} = \expval{r}_{ED} - \expval{r}_{RG}$ is shown in the inset -- adapted from \cite{Aramthottil24}, @(2024) the American Physical Society.}
    \label{fig:bond_disorder}
\end{figure}

Since the disorder is not coupled to onsite operators, the resulting localization is expected to differ significantly from the cases discussed in the preceding sections. As demonstrated in \cite{Aramthottil24}, the description in terms of Local Integrals of Motion (LIOMs) is no longer valid for such systems; instead, a more accurate phenomenological picture is provided by the Real-Space Renormalization Group for excited states (RSRG-X) \cite{Pekker14}. This iterative procedure can be used to asymptotically reproduce the excited eigenstates of the model in the limit of sufficiently large disorder. Fig.~\ref{fig:bond_disorder}(a) compares the ED energy levels with RSRG-X predictions for various disorder strengths. Another peculiar property of Eq.(\ref{eq:bond_disorder_hamiltonian}) is that the half-chain entanglement entropy distributions are peaked around integer values (a feature unparalleled in standard MBL), as shown in Fig.~\ref{fig:bond_disorder}(b).  This behavior is explained by the eigenstate structure, which consists of triplet or singlet states on specific bonds (nicely predicted with RSRG-X). Lastly, bond-disordered models exhibit sub-Poissonian level spacing statistics for a sufficiently large disorder, which is also atypical for standard MBL. A comparison of the gap ratio calculated via RSRG-X and ED is plotted in Fig.~\ref{fig:bond_disorder}(c).

The absence of onsite fields results in the conservation of  $\mathbb{Z}_2$ spin-flip parity symmetry. This symmetry is crucial for investigating various localization-protected orders in excited states \cite{Huse13}, such as Symmetry-Protected Topological (SPT) order \cite{Chandran14, Parameswaran18}, making these models natural candidates for such studies. Crucially, numerical simulations of small system sizes reveal that a portion of the excited spectrum in bond-disordered spin chains exhibits spin-glass order \cite{Parisi83}. Given that spin-glass and SPT orders are generally mutually exclusive \cite{Vasseur16}, the interplay between these competing tendencies represents a compelling property of this class of models \cite{Prodius25}.

\section{Quantum Sun\label{sec:qs}}

The avalanche mechanism~\cite{DeRoeck17, Luitz17} poses a significant challenge to the stability of MBL in the thermodynamic limit. This argument suggests that Griffiths regions of effectively weaker disorder, a natural consequence of random uncorrelated disorder (see Sec.~\ref{sec:qp}) can grow uncontrollably, thermalizing an infinite chain. These concerns inspired the search for effective models describing the interaction between such a thermal region and a localized chain~\cite{Potirniche18,Crowley20,Brighi22l,Suntajs22,Sierant23,Falcao23,Szoldra24,Colmenarez24,Pawlik24}, with the quantum sun model~\cite{Suntajs22,Pawlik24} being a prominent example.
The model is defined by:
\begin{equation}
    \hat{H}_{\mathrm{c}}=R_{\mathrm{c}}\otimes\mathds{1}+\sum\limits_{j=1}^{L}\alpha^{u_{j}}(\hat{S}^{x}_{n_{j}} \hat{S}^{x}_{j}+\hat{S}^{y}_{n_{j}} \hat{S}^{y}_{j})+\sum\limits_{j=1}^L h_j\hat{S}^z_j,\label{eq:sun}
\end{equation}
and consists of two components: the thermal bath of $N$ spins, called the sun, described by a matrix from the GOE with imposed U(1) symmetry, $R_{\rm c}$, and a chain of $L$ localized non-interacting spins, subject only to a disordered magnetic field $h_j\in[1/2,3/2]$ chosen from a uniform distribution independently for each spin. Due to the avalanche mechanism, even a small bath can thermalize the entire localized chain, so the thermodynamic limit of the model is defined by taking fixed and finite $N$.
Each spin $j$ of the chain interacts with a randomly chosen spin of the bath $n_j$ with an interaction strength that decays exponentially with the distance between the given spin and the sun, $u_j\approx j$,  with an exponent $\alpha$. Small randomness is introduced into the positions of spins for numerical reasons~\cite{Suntajs22,Pawlik24}, but this is irrelevant for the physics of the model. Both energy and total spin projection on the $z$ axis are conserved, the latter being especially relevant when the model is mapped to fermions through the Jordan-Wigner transform, as it corresponds to the conservation of the number of particles.

\begin{figure}[ht]
    \centerline{%
        \includegraphics[width=\linewidth]{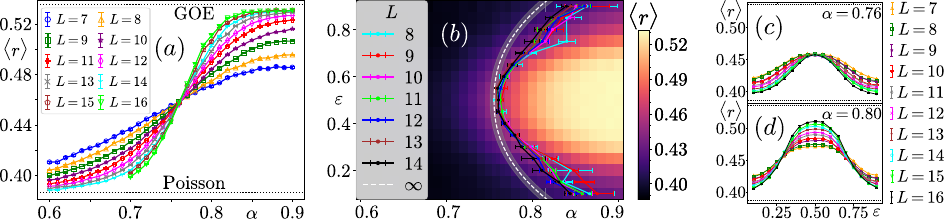}}
    \caption{{\it Ergodicity breaking transition in the U(1) symmetric quantum sun model}. (a) Average gap ratio $\expval{r}$ for various system sizes $L$ and interactions $\alpha$ for the energies in the center of the spectrum $\varepsilon=0.5$. (b) Phase diagram of $\alpha$ against $\varepsilon$, with mobility edge highlighted in white dotted line. Solid lines correspond to the estimate of the ergodicity breaking transition for a given system size $L$. (c, d) Average gap ratio $r$ as a function of $\varepsilon$, showing criticality in the center of the spectrum for $\alpha=0.76$ and ergodicity in the spectral bulk at $\alpha=0.8$ -- adapted from \cite{Pawlik24}, @(2024) the American Physical Society.}
    \label{fig:sun}
\end{figure}

This model is interesting because of its surprisingly small finite-size effects, which plague other MBL models. Figure~\ref{fig:sun}(a) shows the ergodicity breaking phase transition of the U(1) quantum sun model: for small $\alpha$ the system approaches the integrable Poisson level statistics, while above the critical $\alpha\approx0.76$ the system approaches the ergodic behavior of the GOE. The transition between the two behaviors is extremely sharp; curves for different system sizes can be seen to cross almost at a single point somewhere around the critical value of $\alpha$. A quantitative study reveals that some finite-size effects are present, but nonetheless their small magnitude enables one to pinpoint the crossover between localized and ergodic phases with extreme accuracy.

This opens a path to studying different phenomena on the boundary of those phases, most notably the many-body mobility edge: the presence of an energetic boundary separating localized and delocalized states. The mobility edge is a well-established phenomenon for non-interacting systems~\cite{Delande14}, but its status for interacting systems is unclear. Numerical evidence points towards its existence in many-body models~\cite{Luitz15,Chanda20m}, yet some arguments against the possibility of many-body mobility edges in the thermodynamic limit have also been raised~\cite{DeRoeck16}. The quantum sun model, with its well-controlled numerics, offers important insight into this question by accurately providing the dependence of the transition point on energy density $\varepsilon$, see Fig.~\ref{fig:sun}(b). Scaling the results to the thermodynamic limit reveals the existence of a mobility edge, which is in agreement with analytical arguments based on the density of states~\cite{Pawlik24}. Thus, in the ergodic phase only the bulk of the spectrum is fully ergodic, whereas the states at the edges exhibit localized behavior, compare Fig.~\ref{fig:sun}(c) and~\ref{fig:sun}(d).

\section{Conclusions\label{sec:con}}

In this overview, we have explored the multifaceted nature of Many-Body Localization (MBL), emphasizing that the mechanisms of ergodicity breaking extend far beyond the paradigmatic models of random onsite disorder. While standard MBL frameworks rely on uncorrelated onsite fields to induce an ETH-violating phase, we have demonstrated that alternative disorder configurations, especially the quasiperiodic potentials and bond-disordered couplings, offer new physical insights and distinct features.

In particular, quasiperiodic systems provide a cleaner lens for studying the localized phase by suppressing rare-region effects, while bond-disordered spin chains feature a completely different underlying mechanism of localization. Furthermore, we have discussed how the Quantum Sun model serves as a controlled environment for investigating the interaction between a localized chain and a thermal bath. This model remains an essential tool for addressing the fundamental questions surrounding the long-term stability of MBL against thermal inclusions and the potential for an avalanche-driven breakdown of localization in the thermodynamic limit.

In summary, the transition from ergodicity to localization is not a monolithic phenomenon. By moving beyond standard models - especially through research inspired by modern experimental platforms such as Rydberg atom tweezer arrays and optical lattices - we are significantly closer to a comprehensive understanding of ergodicity-breaking phenomena in many-body quantum systems.

\section{Acknowledgements}

We wish professor Jozef Spa\l{}ek many more years of scientific activity. We acknowledge fruitful collaboration on the related topics  with late Dominique Delande and Piotr Sierant. Interesting interactions with Adith Sai Aramthottil, Nicolas Laflorencie, Maciej Lewenstein, Marcin Mierzejewski, and Lev Vidmar are also acknowledged. This work was funded by the National Science Centre, Poland, under the OPUS call within the WEAVE program 2021/43/I/ST3/01142. The research of M.P. and J. Z. was carried out and financed within the framework of the second Swiss Contribution MAPS.
%uncomment the following lines to place a figure
%\begin{figure}[htb]
%\centerline{%
%\includegraphics[width=12.5cm]{Fig1}}
%\caption{Plot of ...}
%\label{Fig:F2H}
%\end{figure}

%\bibliographystyle{IEEEtran}
%\bibliographystyle{ieeetr}
%\bibliography{ref.bib}

\end{document}